\RequirePackage{etex}
\documentclass[letter]{aa}

\usepackage{graphicx}
\usepackage{amsmath}
\usepackage{txfonts}
\usepackage{xspace}
\usepackage{booktabs}
\usepackage{threeparttable}

\def\HI{H {\sc i}\xspace}

\DeclareRobustCommand{\VAN}[3]{#2}
\let\VANthebibliography\thebibliography
\def\thebibliography{\DeclareRobustCommand{\VAN}[3]{##3}\VANthebibliography}

\begin{document}

   \title{A RELHIC twin candidate near the galaxy M51}

   \author{Qingze Chen\inst{1,2}\corrauth{chenqz@bao.ac.cn}
          \and Jie Wang\inst{1,2,3}\corrauth{jie.wang@nao.cas.cn}
          \and Yingjie Jing\inst{1}\corrauth{jyj@nao.cas.cn}
          \and Chen Xu\inst{1,2}
          \and Tiantian Liang\inst{4,5}
          \and Zerui Liu\inst{1,2}
          \and Zhipeng Hou\inst{1,2}
          \and Yiwei Xu\inst{1,2}
          \and Cheng Cheng\inst{1}
          \and Wei Du \inst{1}
          }

    \institute{National Astronomical Observatories, Chinese Academy of Sciences Beijing 100101, China
    \and 
    School of Astronomy and Space Science, University of Chinese Academy of Sciences, Beijing 100049, China
    \and
    Institute for Frontiers in Astronomy and Astrophysics, Beijing Normal University, Beijing 102206, China
    \and
    Department of Mathematics and Physics, North China Electric Power University, Baoding 071003, China
    \and
    Hebei Key Laboratory of Physics and Energy Technology, North China Electric Power University, Baoding 071003, China
    }

   \date{Received XXX; accepted XXX}

   \abstract
   {We report the discovery of a pair of \HI\ clouds in the vicinity of M51 (NGC~5194) using the FAST telescope. These clouds, which share comparable properties and do not display any optical counterpart, are potential candidates for reionisation-limited \HI\ cloud (RELHIC) modelling.}
   {We characterised this newly discovered pair and assessed whether their physical characteristics align with predictions from the RELHIC model.}
   {We performed a systematic search for compact \HI\ sources in the deep \HI\ observations from the FAST Extended-Atlas-of-Selected-Targets Survey (FEASTS) using \textsc{SoFiA}. The resulting catalogue was then cross-matched with the DESI Legacy Imaging Surveys to remove all objects with optical counterparts. Candidate RELHICs were subsequently modelled as hydrostatic \HI\ structures residing in NFW halos and assessed using TNG50 simulations.}
   {We report the detection of two \HI\ clouds (Cloud~S and Cloud~N) located in the outer regions of the M51 system, at projected separations of 70--90~kpc from its centre. Each cloud has an \HI\ mass of $\sim10^{6.5}\,\mathrm{M_\odot}$ and a velocity dispersion of $\sim20~\mathrm{km~s^{-1}}$. Both clouds lack any observable optical counterpart down to a $g$-band surface brightness limit of $\sim27.5~\mathrm{mag~arcsec^{-2}}$. Their stellar luminosities have been constrained to below $10^5\,\mathrm{L_\odot}$. The measured \HI\ characteristics are compatible with RELHIC model predictions and based on our fiducial model, they correspond to characteristic host dark matter halo masses of $3.7\pm0.4\times10^9\,\mathrm{M_\odot}$.}
   {
   Cloud~S and Cloud~N are promising, but not definitive, RELHIC candidates. A tidal origin remains a serious alternative with respect to the interacting environment of M51, particularly since both clouds are unresolved by FAST and the fact that Cloud~N might, in fact, display an internal velocity gradient. Their \HI\ masses and line widths are compatible with RELHIC predictions, but future high-resolution interferometric observations will be essential to distinguish among these possibilities.}

   \keywords{galaxies: dwarf -- galaxies: individual: M51 -- galaxies: ISM -- dark matter -- radio lines: galaxies}

   \maketitle
   \nolinenumbers

\section{Introduction}

In the lambda cold dark matter ($\Lambda$CDM) model, hierarchical galaxy formation may leave some low-mass halos with few or no stars, known as optically dark galaxies \citep{1997MNRAS.292L...5J,2001MNRAS.322..658T}.
Galaxy formation becomes challenging below halo masses of $10^9\,\mathrm{M_\odot}$ and is effectively suppressed below $10^8\,\mathrm{M_\odot}$ \citep{benitez-llambay_detailed_2020}. Simulations show that halos below $10^{9.7}\,\mathrm{M_\odot}$ can be entirely starless \citep{2026MNRAS.547ag385D}, while dedicated studies of dark galaxy populations predict observable \HI-rich systems at the low-mass end \citep{2024ApJ...962..129L,2025arXiv251116726Z}. Identifying such systems provides a critical test of the galaxy formation threshold at the low-mass end of $\Lambda$CDM.

\citet{Benitez-Llambay2017} proposed the reionisation-limited \HI\ cloud (RELHIC) model, characterized by gas in low-mass halos that remains gravitationally bound after reionisation, but does not cool sufficiently to form stars; instead it settles into hydrostatic and thermal equilibrium with the ultraviolet background (UVB) \citep{Benitez-Llambay2017, Rahmati2013}. Finally, the ionising UVB leaves only a centrally concentrated neutral core. RELHICs are predicted to be compact ($\sim1$ kpc), nearly spherical \HI cores with thermally broadened ($\sim20$ km s$^{-1}$) line widths.

\citet{2024ApJ...964...85D} identified `almost dark galaxies' with extremely high \HI-to-stellar mass ratios recorded by ALFALFA. In addition, \citet{2025ApJS..279...38K} compiled a catalogue of dark galaxy candidates, with some genuinely isolated systems certainly figuring among them. 

Recent deep FAST surveys have identified several \HI\ clouds lacking optical counterparts. \citet{2023ApJ...952..130Z} detected a starless \HI\ cloud in the M94 group, Cloud~9, whose physical properties are consistent with RELHIC predictions \citep{2023ApJ...956....1B}. Subsequent VLA observations revealed structural features further supporting a starless $\Lambda$CDM halo interpretation \citep{Benitez-Llambay_2024}, while HST imaging has ruled out a stellar counterpart above $10^{3.5}\,\mathrm{M_\odot}$ at $99.5^{+0.5}_{-8.2}\%$ confidence \citep{2025ApJ...993L..55A}. Cloud~9 is thus the most compelling existing starless halo candidate. Another object, J0139+4328, initially appeared isolated \citep{2023ApJ...944L..40Xu}, but was later found to possess a faint optical counterpart \citep{2026arXiv260112513S,2026A&A...705L...9M}. \citet{2025SciA...11S4057L} reported a compact \HI\ clump ACG185.0$-$11.5 as a potential Local Group dark galaxy. 

\begin{figure}
        \centering
        \includegraphics[width=\columnwidth ]{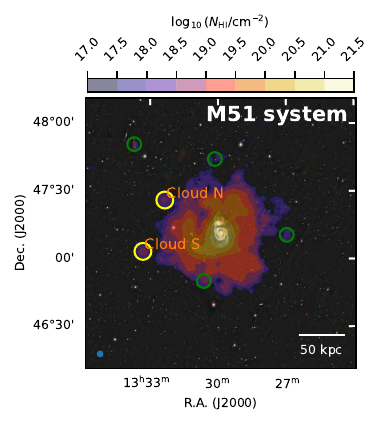}
    \caption{\HI\ distribution in the vicinity of M51 overlaid on a Pan-STARRS1 DR1 optical image. The yellow circles indicate the locations of the two RELHIC candidates: Cloud~N and Cloud~S. The colour shows the \HI\ column density (moment 0) map. The green circles indicate other \HI\ components identified by SoFiA. The blue dot in the bottom-left corner represents the $3.24\arcmin$ FAST beam (FWHM) at 1.42~GHz. A physical scale bar of 50~kpc (assuming a distance of 8~Mpc) is provided in the bottom-right corner.}
    \label{fig:moments}
\end{figure}

In the present work, we report two \HI\ clouds (Cloud~S and Cloud~N, Fig.~\ref{fig:moments}) at projected distances of 70--90~kpc from the centre of the M51 system. These clouds are characterised by \HI\ masses of $\sim10^{6.5}\,\mathrm{M_\odot}$ and velocity dispersions of $\sim20$~km~s$^{-1}$.

\section{FAST \HI\ observations}

The \HI\ data for this study were obtained from the FAST Extended-Atlas-of-Selected-Targets Survey (FEASTS; \citealt{2025ApJ...980...25W}). FEASTS targets the \HI-richest galaxies in the Local Volume, aimed at mapping extended and low-surface-density \HI\ features that are typically below the detection limits of current interferometric observations. The survey utilises the FAST 19-beam receiver in a multibeam on-the-fly (OTF) mapping mode, with each field covered at least six times to minimise contamination from radio frequency interference (RFI) and other artefacts. The resulting data cubes provide a spatial resolution of $3.24\arcmin$ and a velocity resolution of $1.61~\mathrm{km~s^{-1}}$. Notably, the FEASTS moment-0 maps reach a typical $3\sigma$ column density sensitivity of $5 \times 10^{17}~\mathrm{cm^{-2}}$ (assuming a $20~\mathrm{km~s^{-1}}$ line width), making it uniquely suited for detecting the faint neutral cores of RELHICs.

Motivated by the theoretical predictions for RELHICs, we searched the 55-galaxy FEASTS sample for H\,I clouds that satisfied the following criteria:

(i) no optical counterpart in the DESI Legacy Imaging Surveys source catalogue;

(ii) compact and approximately regular H\,I morphology, inconsistent with obvious tidal debris;

(iii) velocity dispersions of order $\sim20$ km s$^{-1}$, as expected for thermally broadened neutral gas in RELHICs.

We identified all \HI sources with \textsc{SoFiA} \citep{Serra_2015}  and rejected all objects failing any criterion above. Among the entire FEASTS sample, only Cloud N and Cloud S in the M51 field satisfied all selection requirements.

\begin{table}
\centering
\newsavebox{\tablebox}
\begin{lrbox}{\tablebox}
\begin{threeparttable}
\begin{tabular}{ccc}
    \toprule
         & Cloud S & Cloud N \\
         \midrule
        R.A. &13h33m15.16s  &13h32m19.22s\\
        Dec. &47d03m29.2s  & 47d26m19.4s\\
        $\log(M_{\mathrm{HI}}/\mathrm{M_\odot})$ & 6.60 &6.53 \\
        Barycentric velocity & 407.2 $\mathrm{km~s^{-1}}$ & 486.2 $\mathrm{km~s^{-1}}$\\
        $W_{50}$ & 21.63 $\mathrm{km~s^{-1}}$ & 29.33 $\mathrm{km~s^{-1}}$\\
        \bottomrule
    \end{tabular}
\end{threeparttable}
\end{lrbox}
\caption{Basic properties of the two optically dark \HI\ clouds.}
\label{tab:1}
\vspace{0.5em}
\scalebox{0.9}{\usebox{\tablebox}}
\end{table}

\section{Results}
\subsection{The two starless \HI\ clouds}
In the output from \textsc{SoFiA}, seven separate \HI\ regions were identified across the data: one large \HI\ region associated with the main body of M51 and six smaller \HI\ clouds. Among the six small clouds, one lies near the edge of the field and has an obvious optical counterpart in the DESI Legacy Imaging Surveys. Another cloud is projected within the extended \HI\ envelope of M51 and is therefore unlikely to represent an isolated system.

The remaining four isolated clouds were further examined using their beam-convolved moment-0 morphologies. We measured the flux-weighted axis ratio ($b/a$) from the second moments of the \HI\ distribution. Two rejected clouds are more elongated, with $b/a=0.50$ and $0.43$, while Cloud~S and Cloud~N have larger apparent axis ratios of $b/a=0.80$ and $0.95$, respectively. These values are used only as beam-convolved morphological descriptors for source selection, since the intrinsic cloud shapes remained unresolved by FAST.

We identified two distinct \HI\ clouds without detectable optical counterparts in the vicinity of M51, a well-known interacting system. These clouds are spatially detached from the main \HI\ envelope of the host galaxy.
Adopting a distance of 8~Mpc to M51 \citep{Bose_Kumar_2014} and a heliocentric velocity of $V_{\rm helio} \approx 460~\mathrm{km~s^{-1}}$, we find that the two clouds (hereafter Cloud~S and Cloud~N) lie at projected distances of 70--90~kpc from the centre of M51. Their physical properties are summarised in Table~\ref{tab:1}.

The intrinsic \HI\ column density profile predicted by the RELHIC model (see Appendix~\ref{sec:relhic_model} for details of the hydrostatic equilibrium modelling) was convolved with a circular Gaussian beam with a full width at half maximum (FWHM) of $3.24\arcmin$, corresponding to the angular resolution of the FAST \HI\ observations. Owing to the compact intrinsic size of RELHICs relative to the FAST beam, these objects generally appear as point-like sources.

\begin{figure}
        \includegraphics[width=\columnwidth]{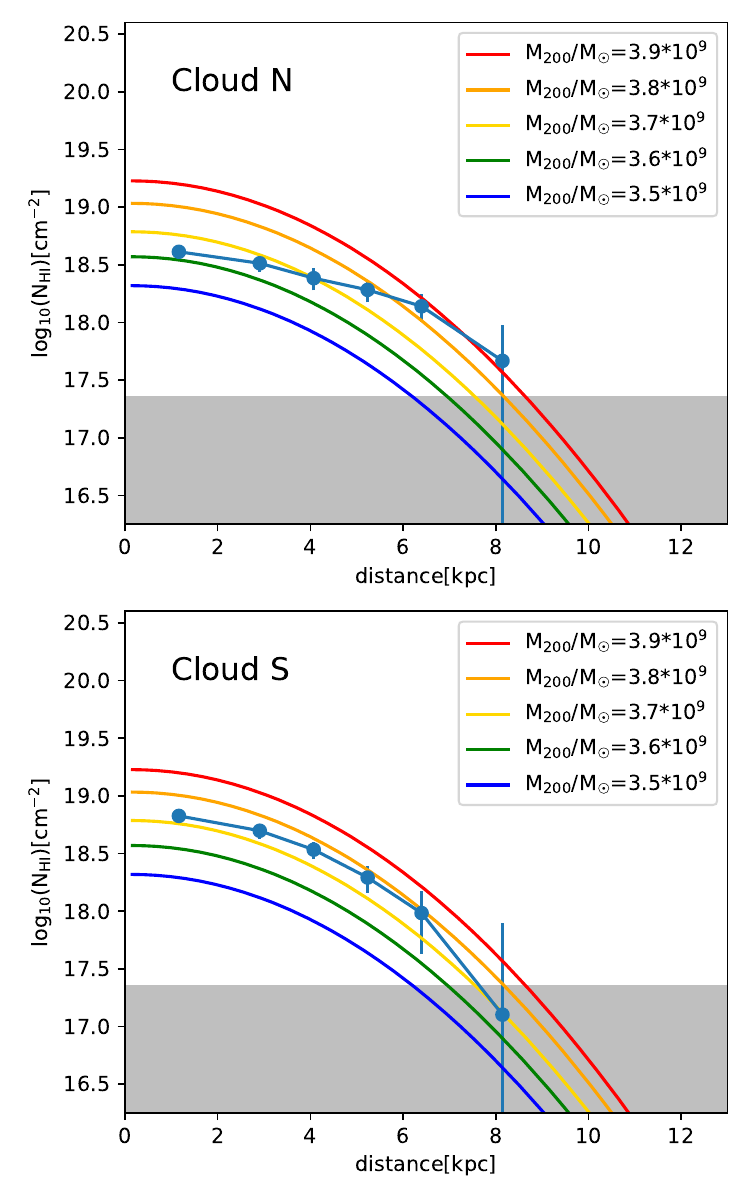}
    \caption{Radial \HI\ column density profiles for the two RELHIC candidates. Blue circles with error bars represent the observed profiles measured from the FAST data. The solid curves denote a series of theoretical RELHIC models calculated for the distance of M51 (8~Mpc) and convolved with the FAST Gaussian beam. The different curves correspond to varying halo masses, with the best-fit model suggesting $M_{200} \approx 3.7 \times 10^9~\mathrm{M_\odot}$.}
    \label{fig:profiles}
\end{figure}

We compared the convolved \HI\ radial distributions with the observed \HI\ profiles of Cloud~N and Cloud~S. As shown in Fig.~\ref{fig:profiles}, both clouds exhibit \HI\ profiles consistent with the RELHIC model prediction. By comparing the observed profiles with the beam-convolved RELHIC predictions, we estimate that the halo mass of both clouds is approximately $3.7 \times 10^9\,\mathrm{M_\odot}$ (Appendix~\ref{sec:relhic_model}). The \HI\ column density ($n_{\mathrm{HI}}$) is around $10^{19}~\mathrm{cm^{-2}}$ after being convolved with the FAST beam; the intrinsic $n_{\mathrm{HI}}$ is expected to be two to three orders of magnitude higher. This implies that the intrinsic gas structure could be detectable with higher-resolution interferometric observations.
To assess the reliability of this mass estimation approach, we further compared the RELHIC-derived halo masses with starless subhalos in the TNG50 cosmological simulation. The comparison shows that the inferred masses are generally consistent with the true halo masses within a factor of 2 (Appendix~\ref{sec:tng_validation}).

\subsection{Optical data and detection limits}

To confirm the starless nature of our candidates, we utilised multi-band optical imaging from the Dark Energy Spectroscopic Instrument (DESI) Legacy Imaging Surveys \citep{Dey2019} and Pan-STARRS1 (PS1) \citep{Flewelling2020}. While PS1 provides superior sky coverage, the Legacy Survey, specifically the BASS and MzLS footprints covering the M51 region, offers significantly deeper imaging. For the M51 field, we adopted the BASS $g$-band median $5\sigma$ point-source depth of 24.3~mag as a baseline.

We marked the extragalactic sources within the field of view in Fig.~\ref{fig:zoom in}. These sources were selected based on either photometric or spectroscopic redshift measurements with $z < 0.2$.
None of the identified extragalactic sources has a redshift consistent with that of the \HI\ clouds, which corresponds to $z \approx 0.0013$. This lack of redshift coincidence indicates that there are no known background or foreground galaxies directly associated with the detected \HI\ clouds.

To quantify our sensitivity to low-surface-brightness stellar components, we calculated surface brightness limits using the \texttt{sbcontrast} code \citep{Keim_2022}. The resulting $3\sigma$ surface brightness limits, measured on a $10\arcsec \times 10\arcsec$ scale, are 27.7, 27.1, and 26.5~$\mathrm{mag~arcsec^{-2}}$ in the $g$, $r$, and $z$ bands, respectively.

\begin{figure}
    \centering
    \includegraphics[width =\linewidth]{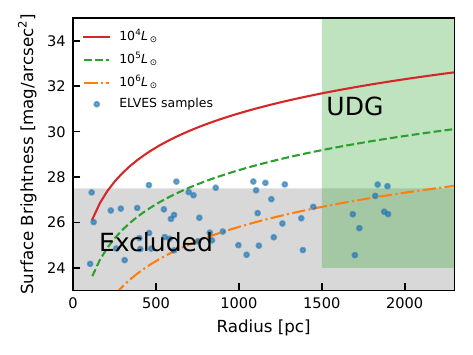}
    \caption{Constraints on the stellar luminosity and structural properties of the RELHIC candidates. Blue dots represent the dwarf satellites from the ELVES survey \citep{Carlsten_2022} for comparison. The grey shaded region indicates the parameter space excluded by our $3\sigma$ surface brightness limit of $\mu_g \approx 27.5~\mathrm{mag~arcsec^{-2}}$. The green shaded area defines the region occupied by ultra-diffuse galaxies (UDGs). The three curves illustrate the relationship between surface brightness and physical radius for galaxies with fixed total luminosities of $10^4$, $10^5$, and $10^6~\mathrm{L_\odot}$ at a distance of 8~Mpc. As the radius increases, the surface brightness decreases, moving the candidates further into the undetected regime. The luminosity of any optical counterpart is unlikely to exceed $10^6~\mathrm{L_\odot}$, and is most likely below $10^5~\mathrm{L_\odot}$.}
    \label{fig:OCP}
\end{figure}

Despite the robust \HI\ detections, no stellar counterparts were identified in the deep BASS $g$-band imaging. We therefore adopted a conservative $3\sigma$ surface brightness limit of $\sim 27.5~\mathrm{mag~arcsec^{-2}}$ in the $g$ band at a $10\arcsec$ scale for the M51 field. To quantify the corresponding luminosity upper limit ($L_{\rm lim}$), we adopted a characteristic physical radius motivated by Local Volume dwarf galaxies from the ELVES survey \citep{Carlsten_2022}. As illustrated in Fig.~\ref{fig:OCP}, the surface brightness of a galaxy with a given luminosity decreases as its effective radius increases. Assuming a typical dwarf satellite effective radius of $r_{\rm e} \sim 500$~pc, we derived a luminosity limit of $L_{\rm lim} \sim 10^5\,\mathrm{L_\odot}$. 

This value corresponds to an exceptionally high \HI\ mass-to-light ratio of $M_{\mathrm{HI}}/L \gtrsim 30\,\mathrm{M_\odot/L_\odot}$, which supports the interpretation of these objects as dark galaxy candidates. Accordingly, if a potential stellar component is found to be more spatially extended, the upper limit on the luminosity would be higher.

\section{Summary and discussion}
In this work, we report the discovery of two starless \HI\ cloud candidates in the outskirts of the M51 system. The two clouds share comparable \HI\ masses, velocity dispersions, and projected distances from M51. They are both compatible with the RELHIC framework. Our primary findings are summarised below.

\begin{enumerate}
\item The observed \HI\ masses ($\sim 10^{6.5}\,\mathrm{M_\odot}$) and velocity dispersions ($\sim 20\,\mathrm{km\,s^{-1}}$) are compatible with the theoretical predictions of the RELHIC model.
\item By applying the hydrostatic equilibrium condition within the RELHIC framework, we were able to estimate the characteristic halo masses of these candidates as $M_{200}\simeq3.7\pm0.4\times10^9\,\mathrm{M_\odot}$.
\end{enumerate}

Given that M51 is a well-known interacting system, a primary concern is whether Cloud~S and Cloud~N could be tidal debris: pure gas clouds stripped from the host galaxy during interactions. We acknowledge this possibility and we do not claim that the present FAST data has the capacity to unambiguously distinguish between the tidal and RELHIC scenarios. Nevertheless, several observed properties remain consistent with the RELHIC interpretation.

Tidal debris is typically manifested as large-scale structures with broad velocity dispersions, such as the south-eastern tidal tail reported by \citet{2023MNRAS.521.2719Y}. The total mass of the M51 system is approximately $3\times10^{11}\,\mathrm{M_\odot}$, based on the simulations of \citet{2010MNRAS.403..625D}. For a dark matter halo with a mass of $3.7\times10^9\,\mathrm{M_\odot}$ located at a galactocentric distance of 100~kpc from M51, the corresponding tidal radius is estimated to be $\sim$19~kpc. Therefore, an \HI\ cloud embedded within such a dark matter halo could remain gravitationally bound and survive tidal disruption at this distance.

The narrow line widths expected for thermally broadened gas, together with the high \HI-to-light ratios, are more naturally expected for RELHIC candidates. While we cannot entirely rule out a tidal origin without access to higher resolution data, the agreement with $\Lambda$CDM-based RELHIC predictions makes them worthy of follow-up as starless dark matter halo candidates. Since FAST cannot resolve the small-sized \HI\ clouds, interferometric observations are needed to obtain more accurate measurements of the intrinsic shapes and radii of these clouds.

While current surface brightness limits suggest a stellar mass below $10^5~\mathrm{M_\odot}$, deep optical imaging (e.g., with HST) could allow us to resolve individual bright stars at a distance of 8~Mpc. By conducting a statistical analysis of the colour-magnitude diagram for point sources within the \HI\ cloud region compared to adjacent control fields, we could search for a localised stellar excess. This method would either lead to the detection of a hidden, extremely low-surface-brightness stellar component or place an even more stringent upper limit on star formation efficiency in low-mass dark matter halos.

\begin{acknowledgements}
This work was supported by the China National Key Program for Science and Technology Research and Development of China (No. 2022YFA1602901). This work made use of data from FAST (Five-hundred-metre Aperture Spherical radio Telescope; \url{https://cstr.cn/31116.02.FAST}). FAST is a Chinese national mega-science facility, operated by the National Astronomical Observatories, Chinese Academy of Sciences. We are grateful to the FEASTS survey for providing data. We would also like to thank Julio Navarro, Carlos Frenk, and Simon White for their insightful and fruitful discussions. This work made use of data from the DESI Legacy Imaging Surveys and Pan-STARRS1. This work is supported by the Fundamental Research Funds for the Central Universities (2026MS217).
\end{acknowledgements}

\bibliographystyle{bibtex/aa}
\bibliography{example}

@ARTICLE{2024JOSS....9.6860O,
    author = {{Oman}, Kyle A.},
    title = "{MARTINI: Mock Array Radio Telescope Interferometry of the Neutral ISM}",
    journal = {The Journal of Open Source Software},
    year = 2024,
    month = jun,
    volume = {9},
    number = {98},
    eid = {6860},
    pages = {6860},
    doi = {10.21105/joss.06860},
    adsurl = {https://ui.adsabs.harvard.edu/abs/2024JOSS....9.6860O}
}

@ARTICLE{2016MNRAS.460.1214L,
       author = {{Ludlow}, Aaron D. and {Bose}, Sownak and {Angulo}, Ra{\'u}l E. and {Wang}, Lan and {Hellwing}, Wojciech A. and {Navarro}, Julio F. and {Cole}, Shaun and {Frenk}, Carlos S.},
        title = "{The mass-concentration-redshift relation of cold and warm dark matter haloes}",
      journal = {\mnras},
         year = 2016,
        month = aug,
       volume = {460},
       number = {2},
        pages = {1214-1232},
          doi = {10.1093/mnras/stw1046},
archivePrefix = {arXiv},
       eprint = {1601.02624},
 primaryClass = {astro-ph.CO},
       adsurl = {https://ui.adsabs.harvard.edu/abs/2016MNRAS.460.1214L}
}

@ARTICLE{2025ApJ...980...25W,
       author = {{Wang}, Jing and {Yang}, Dong and {Lin}, Xuchen and {Huang}, Qifeng and {Qu}, Zhijie and {Chen}, Hsiao-wen and {Guo}, Hong and {Ho}, Luis C. and {Jiang}, Peng and {Liang}, Zezhong and {P{\'e}roux}, C{\'e}line and {Staveley-Smith}, Lister and {Weng}, Simon},
        title = "{FEASTS: Radial Distribution of H I Surface Densities Down to 0.01 M$_{{\ensuremath{\odot}}}$ pc$^{{\ensuremath{-}}2}$ of 35 Nearby Galaxies}",
      journal = {\apj},
         year = 2025,
        month = feb,
       volume = {980},
       number = {1},
          eid = {25},
        pages = {25},
          doi = {10.3847/1538-4357/ada95a},
archivePrefix = {arXiv},
       eprint = {2501.01289},
 primaryClass = {astro-ph.GA},
       adsurl = {https://ui.adsabs.harvard.edu/abs/2025ApJ...980...25W}
}

@article{Flewelling2020,
   author = {H. A. Flewelling and E. A. Magnier and K. C. Chambers and J. N. Heasley and C. Holmberg and M. E. Huber and W. Sweeney and C. Z. Waters and A. Calamida and S. Casertano and X. Chen and D. Farrow and G. Hasinger and R. Henderson and K. S. Long and N. Metcalfe and G. Narayan and M. A. Nieto-Santisteban and P. Norberg and A. Rest and R. P. Saglia and A. Szalay and A. R. Thakar and J. L. Tonry and J. Valenti and S. Werner and R. White and L. Denneau and P. W. Draper and K. W. Hodapp and R. Jedicke and N. Kaiser and R. P. Kudritzki and P. A. Price and R. J. Wainscoat and S. Chastel and B. McLean and M. Postman and B. Shiao},
   doi = {10.3847/1538-4365/abb82d},
   issn = {0067-0049},
   issue = {1},
   journal = {The Astrophysical Journal Supplement Series},
   month = {11},
   pages = {7},
   title = {The Pan-STARRS1 Database and Data Products},
   volume = {251},
   year = {2020},
}

@article{Carlsten_2022,
doi = {10.3847/1538-4357/ac6fd7},
url = {https://dx.doi.org/10.3847/1538-4357/ac6fd7},
year = {2022},
month = {jul},
publisher = {The American Astronomical Society},
volume = {933},
number = {1},
pages = {47},
author = {Scott G. Carlsten and Jenny E. Greene and Rachael L. Beaton and Shany Danieli and Johnny P. Greco},
title = {The Exploration of Local VolumE Satellites (ELVES) Survey: A Nearly Volume-limited Sample of Nearby Dwarf Satellite Systems},
journal = {The Astrophysical Journal}
}

@article{Benitez-Llambay2017,
   author = {Alejandro Benítez-Llambay and Julio F. Navarro and Carlos S. Frenk and Till Sawala and Kyle Oman and Azadeh Fattahi and Matthieu Schaller and Joop Schaye and Robert A. Crain and Tom Theuns},
   doi = {10.1093/mnras/stw2982},
   issn = {13652966},
   issue = {4},
   journal = {Monthly Notices of the Royal Astronomical Society},
   month = {3},
   pages = {3913-3926},
   title = {The properties of 'dark' $\Lambda$CDM haloes in the Local Group},
   volume = {465},
   url = {http://arxiv.org/abs/1609.01301 http://dx.doi.org/10.1093/mnras/stw2982 https://academic.oup.com/mnras/article-lookup/doi/10.1093/mnras/stw2982},
   year = {2017},
}

@article{Rahmati2013,
   author = {Alireza Rahmati and Andreas H. Pawlik and Milan Raicevic and Joop Schaye},
   doi = {10.1093/mnras/stt066},
   issn = {1365-2966},
   issue = {3},
   journal = {Monthly Notices of the Royal Astronomical Society},
   month = {4},
   pages = {2427-2445},
   title = {On the evolution of the Hi column density distribution in cosmological simulations},
   volume = {430},
   url = {http://academic.oup.com/mnras/article/430/3/2427/1751889/On-the-evolution-of-the-Hi-column-density},
   year = {2013},
}

@article{Serra_2015,
   title={SoFiA: a flexible source finder for 3D spectral line data},
   volume={448},
   ISSN={0035-8711},
   url={http://dx.doi.org/10.1093/mnras/stv079},
   DOI={10.1093/mnras/stv079},
   number={2},
   journal={Monthly Notices of the Royal Astronomical Society},
   publisher={Oxford University Press (OUP)},
   author={Serra, Paolo and Westmeier, Tobias and Giese, Nadine and Jurek, Russell and Flöer, Lars and Popping, Attila and Winkel, Benjamin and van der Hulst, Thijs and Meyer, Martin and Koribalski, Bärbel S. and Staveley-Smith, Lister and Courtois, Hélène},
   year={2015},
   month=feb, pages={1922–1929} }

@article{Dey2019,
   author = {Arjun Dey and David J. Schlegel and Dustin Lang and Robert Blum and Kaylan Burleigh and Xiaohui Fan and Joseph R. Findlay and Doug Finkbeiner and David Herrera and Stéphanie Juneau and Martin Landriau and Michael Levi and Ian McGreer and Aaron Meisner and Adam D. Myers and John Moustakas and Peter Nugent and Anna Patej and Edward F. Schlafly and Alistair R. Walker and Francisco Valdes and Benjamin A. Weaver and Christophe Yèche and Hu Zou and Xu Zhou and Behzad Abareshi and T. M. C. Abbott and Bela Abolfathi and C. Aguilera and Shadab Alam and Lori Allen and A. Alvarez and James Annis and Behzad Ansarinejad and Marie Aubert and Jacqueline Beechert and Eric F. Bell and Segev Y. BenZvi and Florian Beutler and Richard M. Bielby and Adam S. Bolton and César Briceño and Elizabeth J. Buckley-Geer and Karen Butler and Annalisa Calamida and Raymond G. Carlberg and Paul Carter and Ricard Casas and Francisco J. Castander and Yumi Choi and Johan Comparat and Elena Cukanovaite and Timothée Delubac and Kaitlin DeVries and Sharmila Dey and Govinda Dhungana and Mark Dickinson and Zhejie Ding and John B. Donaldson and Yutong Duan and Christopher J. Duckworth and Sarah Eftekharzadeh and Daniel J. Eisenstein and Thomas Etourneau and Parker A. Fagrelius and Jay Farihi and Mike Fitzpatrick and Andreu Font-Ribera and Leah Fulmer and Boris T. Gänsicke and Enrique Gaztanaga and Koshy George and David W. Gerdes and Satya Gontcho A Gontcho and Claudio Gorgoni and Gregory Green and Julien Guy and Diane Harmer and M. Hernandez and Klaus Honscheid and Lijuan (Wendy) Huang and David J. James and Buell T. Jannuzi and Linhua Jiang and Richard Joyce and Armin Karcher and Sonia Karkar and Robert Kehoe and Jean-Paul, Kneib and Andrea Kueter-Young and Ting-Wen Lan and Tod R. Lauer and Laurent Le Guillou and Auguste Le Van Suu and Jae Hyeon Lee and Michael Lesser and Laurence Perreault Levasseur and Ting S. Li and Justin L. Mann and Robert Marshall and C. E. Martínez-Vázquez and Paul Martini and Hélion du Mas des Bourboux and Sean McManus and Tobias Gabriel Meier and Brice Ménard and Nigel Metcalfe and Andrea Muñoz-Gutiérrez and Joan Najita and Kevin Napier and Gautham Narayan and Jeffrey A. Newman and Jundan Nie and Brian Nord and Dara J. Norman and Knut A. G. Olsen and Anthony Paat and Nathalie Palanque-Delabrouille and Xiyan Peng and Claire L. Poppett and Megan R. Poremba and Abhishek Prakash and David Rabinowitz and Anand Raichoor and Mehdi Rezaie and A. N. Robertson and Natalie A. Roe and Ashley J. Ross and Nicholas P. Ross and Gregory Rudnick and Sasha Gaines and Abhijit Saha and F. Javier Sánchez and Elodie Savary and Heidi Schweiker and Adam Scott and Hee-Jong Seo and Huanyuan Shan and David R. Silva and Zachary Slepian and Christian Soto and David Sprayberry and Ryan Staten and Coley M. Stillman and Robert J. Stupak and David L. Summers and Suk Sien Tie and H. Tirado and Mariana Vargas-Magaña and A. Katherina Vivas and Risa H. Wechsler and Doug Williams and Jinyi Yang and Qian Yang and Tolga Yapici and Dennis Zaritsky and A. Zenteno and Kai Zhang and Tianmeng Zhang and Rongpu Zhou and Zhimin Zhou},
   doi = {10.3847/1538-3881/ab089d},
   issn = {0004-6256},
   issue = {5},
   journal = {The Astronomical Journal},
   month = {5},
   pages = {168},
   title = {Overview of the DESI Legacy Imaging Surveys},
   volume = {157},
   year = {2019},
}

@article{benitez-llambay_detailed_2020,
	title = {The detailed structure and the onset of galaxy formation in low-mass gaseous dark matter haloes},
	volume = {498},
	issn = {0035-8711},
	url = {https://doi.org/10.1093/mnras/staa2698},
	doi = {10.1093/mnras/staa2698},
	pages = {4887--4900},
	number = {4},
	journal = {Monthly Notices of the Royal Astronomical Society},
	shortjournal = {Monthly Notices of the Royal Astronomical Society},
	author = {Benitez-Llambay, Alejandro and Frenk, Carlos},
	urldate = {2025-01-12},
	date = {2020-10-10},
         year = {2020},
}

@ARTICLE{2023ApJ...944L..40Xu,
       author = {{Xu}, Jin-Long and {Zhu}, Ming and {Yu}, Naiping and {Zhang}, Chuan-Peng and {Liu}, Xiao-Lan and {Ai}, Mei and {Jiang}, Peng},
        title = "{Discovery of an Isolated Dark Dwarf Galaxy in the Nearby Universe}",
      journal = {\apjl},
         year = 2023,
        month = feb,
       volume = {944},
       number = {2},
          eid = {L40},
        pages = {L40},
          doi = {10.3847/2041-8213/acb932},
archivePrefix = {arXiv},
       eprint = {2302.02646},
 primaryClass = {astro-ph.CO},
       adsurl = {https://ui.adsabs.harvard.edu/abs/2023ApJ...944L..40X}
}

@ARTICLE{2023ApJ...956....1B,
       author = {{Benitez-Llambay}, Alejandro and {Navarro}, Julio F.},
        title = "{Is a Recently Discovered H I Cloud near M94 a Starless Dark Matter Halo?}",
      journal = {\apj},
         year = 2023,
        month = oct,
       volume = {956},
       number = {1},
          eid = {1},
        pages = {1},
          doi = {10.3847/1538-4357/acf767},
archivePrefix = {arXiv},
       eprint = {2309.03253},
 primaryClass = {astro-ph.GA},
       adsurl = {https://ui.adsabs.harvard.edu/abs/2023ApJ...956....1B}
}

@ARTICLE{2023ApJ...952..130Z,
       author = {{Zhou}, Ruilei and {Zhu}, Ming and {Yang}, Yanbin and {Yu}, Haiyang and {Yuan}, Lixia and {Jiang}, Peng and {Xi}, Wenzhe},
        title = "{FAST Reveals New Evidence for M94 as a Merger}",
      journal = {\apj},
         year = 2023,
        month = aug,
       volume = {952},
       number = {2},
          eid = {130},
        pages = {130},
          doi = {10.3847/1538-4357/acdcf5},
archivePrefix = {arXiv},
       eprint = {2306.05080},
 primaryClass = {astro-ph.GA},
       adsurl = {https://ui.adsabs.harvard.edu/abs/2023ApJ...952..130Z}
}

@article{Benitez-Llambay_2024,
doi = {10.3847/1538-4357/ad65d9},
url = {https://dx.doi.org/10.3847/1538-4357/ad65d9},
year = {2024},
month = {sep},
publisher = {The American Astronomical Society},
volume = {973},
number = {1},
pages = {61},
author = {Benítez-Llambay, Alejandro and Dutta, Rajeshwari and Fumagalli, Michele and Navarro, Julio F.},
title = {Examining the Nature of the Starless Dark Matter Halo Candidate Cloud-9 with Very Large Array Observations},
journal = {The Astrophysical Journal}
}

@ARTICLE{2001MNRAS.322..658T,
       author = {{Trentham}, Neil and {M{\"o}ller}, Ole and {Ramirez-Ruiz}, Enrico},
        title = "{Completely dark galaxies: their existence, properties and strategies for finding them}",
      journal = {\mnras},
         year = 2001,
        month = apr,
       volume = {322},
       number = {3},
        pages = {658-668},
          doi = {10.1046/j.1365-8711.2001.04158.x},
archivePrefix = {arXiv},
       eprint = {astro-ph/0010545},
 primaryClass = {astro-ph},
       adsurl = {https://ui.adsabs.harvard.edu/abs/2001MNRAS.322..658T}
}

@ARTICLE{1997MNRAS.292L...5J,
       author = {{Jimenez}, R. and {Heavens}, A.~F. and {Hawkins}, M.~R.~S. and {Padoan}, P.},
        title = "{Dark galaxies, spin bias and gravitational lenses}",
      journal = {\mnras},
         year = 1997,
        month = nov,
       volume = {292},
       number = {1},
        pages = {L5-L10},
          doi = {10.1093/mnras/292.1.L5},
archivePrefix = {arXiv},
       eprint = {astro-ph/9709050},
 primaryClass = {astro-ph},
       adsurl = {https://ui.adsabs.harvard.edu/abs/1997MNRAS.292L...5J}
}

@article{Nelson_2019,
   title={First results from the TNG50 simulation: galactic outflows driven by supernovae and black hole feedback},
   volume={490},
   ISSN={1365-2966},
   url={http://dx.doi.org/10.1093/mnras/stz2306},
   DOI={10.1093/mnras/stz2306},
   number={3},
   journal={Monthly Notices of the Royal Astronomical Society},
   publisher={Oxford University Press (OUP)},
   author={Nelson, Dylan and Pillepich, Annalisa and Springel, Volker and Pakmor, Rüdiger and Weinberger, Rainer and Genel, Shy and Torrey, Paul and Vogelsberger, Mark and Marinacci, Federico and Hernquist, Lars},
   year={2019},
   month=aug, pages={3234–3261} }

@article{Pillepich_2019,
   title={First results from the TNG50 simulation: the evolution of stellar and gaseous discs across cosmic time},
   volume={490},
   ISSN={1365-2966},
   url={http://dx.doi.org/10.1093/mnras/stz2338},
   DOI={10.1093/mnras/stz2338},
   number={3},
   journal={Monthly Notices of the Royal Astronomical Society},
   publisher={Oxford University Press (OUP)},
   author={Pillepich, Annalisa and Nelson, Dylan and Springel, Volker and Pakmor, Rüdiger and Torrey, Paul and Weinberger, Rainer and Vogelsberger, Mark and Marinacci, Federico and Genel, Shy and van der Wel, Arjen and Hernquist, Lars},
   year={2019},
   month=sep, pages={3196–3233} }

@ARTICLE{2026MNRAS.547ag385D,
       author = {{Doppel}, Jessica E. and {Jauzac}, Mathilde and {Lagattuta}, David J. and {Fattahi}, Azadeh and {Mahler}, Guillaume},
        title = "{Tiny galaxies and dark substructures: exploring the 'dark' subhaloes in TNG50}",
      journal = {\mnras},
         year = 2026,
        month = apr,
       volume = {547},
       number = {3},
          eid = {stag385},
        pages = {stag385},
          doi = {10.1093/mnras/stag385},
archivePrefix = {arXiv},
       eprint = {2506.09122},
 primaryClass = {astro-ph.GA},
       adsurl = {https://ui.adsabs.harvard.edu/abs/2026MNRAS.547ag385D}
}

@INPROCEEDINGS{2001cghr.confE..64H,
       author = {{Haardt}, F. and {Madau}, P.},
        title = "{Modelling the UV/X-ray cosmic background with CUBA}",
    booktitle = {Clusters of Galaxies and the High Redshift Universe Observed in X-rays},
         year = 2001,
       editor = {{Neumann}, D.~M. and {Tran}, J.~T.~V.},
        month = jan,
          eid = {64},
        pages = {64},
          doi = {10.48550/arXiv.astro-ph/0106018},
archivePrefix = {arXiv},
       eprint = {astro-ph/0106018},
 primaryClass = {astro-ph},
       adsurl = {https://ui.adsabs.harvard.edu/abs/2001cghr.confE..64H}
}

@article{Bose_Kumar_2014, title={DISTANCE DETERMINATION TO EIGHT GALAXIES USING EXPANDING PHOTOSPHERE METHOD}, volume={782}, url={http://dx.doi.org/10.1088/0004-637X/782/2/98}, DOI={10.1088/0004-637x/782/2/98}, number={2}, journal={The Astrophysical Journal}, publisher={American Astronomical Society}, author={Bose, Subhash and Kumar, Brijesh}, year={2014}, month=feb, pages={98} }

@article{Keim_2022,
   title={Tidal Distortions in NGC1052-DF2 and NGC1052-DF4: Independent Evidence for a Lack of Dark Matter},
   volume={935},
   ISSN={1538-4357},
   url={http://dx.doi.org/10.3847/1538-4357/ac7dab},
   DOI={10.3847/1538-4357/ac7dab},
   number={2},
   journal={The Astrophysical Journal},
   publisher={American Astronomical Society},
   author={Keim, Michael A. and Dokkum, Pieter van and Danieli, Shany and Lokhorst, Deborah and Li , Jiaxuan  and Shen, Zili and Abraham, Roberto and Chen, Seery and Gilhuly, Colleen and Liu , Qing  and Merritt, Allison and Miller, Tim B. and Pasha, Imad and Polzin, Ava},
   year={2022},
   month=aug, pages={160} }

@ARTICLE{2024ApJ...964...85D,
       author = {{Du}, Lin and {Du}, Wei and {Cheng}, Cheng and {Zhu}, Ming and {Yu}, Haiyang and {Wu}, Hong},
        title = "{Almost Optically Dark Galaxies in DECaLS (I): Detection, Optical Properties, and Possible Origins}",
      journal = {\apj},
         year = 2024,
        month = mar,
       volume = {964},
       number = {1},
          eid = {85},
        pages = {85},
          doi = {10.3847/1538-4357/ad234f},
archivePrefix = {arXiv},
       eprint = {2403.12130},
 primaryClass = {astro-ph.GA},
       adsurl = {https://ui.adsabs.harvard.edu/abs/2024ApJ...964...85D}
}

@ARTICLE{2025arXiv251116726Z,
author = {{Zheng}, Haonan and {Jiang}, Fangzhou and {Liao}, Shihong and {Libeskind}, Noam I.},
       author = {{Zheng}, Haonan and {Jiang}, Fangzhou and {Liao}, Shihong and {Libeskind}, Noam I.},
        title = "{HIDES. I. The Population and Diversity of H I-rich Faint Dwarf Galaxies in the HESTIA and Auriga Simulations}",
      journal = {\apj},
         year = 2026,
        month = jun,
       volume = {1004},
       number = {1},
          eid = {79},
        pages = {79},
          doi = {10.3847/1538-4357/ae6b84},
archivePrefix = {arXiv},
       eprint = {2511.16726},
 primaryClass = {astro-ph.GA},
       adsurl = {https://ui.adsabs.harvard.edu/abs/2026ApJ..1004...79Z}
}

@ARTICLE{2024ApJ...962..129L,
author = {{Lee}, Gain and {Hwang}, Ho Seong and {Lee}, Jaehyun and {Shin}, Jihye and {Song}, Hyunmi},
title = "{Understanding the Formation and Evolution of Dark Galaxies in a Simulated Universe}",
journal = {\apj},
year = 2024,
month = feb,
volume = {962},
number = {2},
eid = {129},
pages = {129},
doi = {10.3847/1538-4357/ad1e5d},
archivePrefix = {arXiv},
eprint = {2401.07007},
primaryClass = {[astro-ph.GA](http://astro-ph.ga/)},
adsurl = {https://ui.adsabs.harvard.edu/abs/2024ApJ...962..129L}
}

@ARTICLE{2025ApJS..279...38K,
author = {{Kwon}, Minseong and {Hwang}, Ho Seong and {Kent}, Brian R. and {Yoon}, Ilsang and {Lee}, Gain and {Yoon}, Hyein},
title = "{Searching for Dark Galaxies with H I Detection from the Arecibo Legacy Fast ALFA (ALFALFA) Survey}",
journal = {\apjs},
year = 2025,
month = aug,
volume = {279},
number = {2},
eid = {38},
pages = {38},
doi = {10.3847/1538-4365/ade0b8},
archivePrefix = {arXiv},
eprint = {2506.03678},
primaryClass = {[astro-ph.GA](http://astro-ph.ga/)},
adsurl = {https://ui.adsabs.harvard.edu/abs/2025ApJS..279...38K}
}

@ARTICLE{2025SciA...11S4057L,
       author = {{Liu}, Xiao-Lan and {Xu}, Jin-Long and {Jiang}, Peng and {Zhu}, Ming and {Zhang}, Chuan-Peng and {Yu}, Naiping and {Xu}, Ye and {Guan}, Xin and {Wang}, Jun-Jie},
        title = "{Discovery of a high-velocity cloud of the Milky Way as a potential dark galaxy}",
      journal = {Science Advances},
         year = 2025,
        month = apr,
       volume = {11},
       number = {16},
          eid = {eads4057},
        pages = {eads4057},
          doi = {10.1126/sciadv.ads4057},
archivePrefix = {arXiv},
       eprint = {2504.09419},
 primaryClass = {astro-ph.GA},
       adsurl = {https://ui.adsabs.harvard.edu/abs/2025SciA...11S4057L}
}

@ARTICLE{2026A&A...705L...9M,
       author = {{Mitra{\v{s}}inovi{\'c}}, Ana and {Grozdanovi{\'c}}, Marko and {Lalovi{\'c}}, Ana and {Jovanovi{\'c}}, Milena and {B{\'\i}lek}, Michal and {Pavlov}, Nata{\v{s}}a and {Moiseev}, Alexei V. and {Oparin}, Dmitry V.},
        title = "{Discovery of a galaxy associated with the HI cloud FAST J0139+4328}",
      journal = {\aap},
         year = 2026,
        month = jan,
       volume = {705},
          eid = {L9},
        pages = {L9},
          doi = {10.1051/0004-6361/202558391},
archivePrefix = {arXiv},
       eprint = {2512.24924},
 primaryClass = {astro-ph.GA},
       adsurl = {https://ui.adsabs.harvard.edu/abs/2026A&A...705L...9M}
}

@ARTICLE{2026arXiv260112513S,
       author = {{{\v{S}}iljeg}, Barbara and {Adams}, Elizabeth A.~K. and {Oosterloo}, Tom A. and {Fraternali}, Filippo and {Hess}, Kelley M. and {Xu}, Jin-Long and {Zhu}, Ming},
        title = "{Not so-dark: High resolution H I imaging of J0139+4328 and identification of an optical counterpart}",
      journal = {\aap},
         year = 2026,
        month = mar,
       volume = {708},
          eid = {A40},
        pages = {A40},
          doi = {10.1051/0004-6361/202556900},
archivePrefix = {arXiv},
       eprint = {2601.12513},
 primaryClass = {astro-ph.GA},
       adsurl = {https://ui.adsabs.harvard.edu/abs/2026A&A...708A..40S}
}

@ARTICLE{2010MNRAS.403..625D,
       author = {{Dobbs}, C.~L. and {Theis}, C. and {Pringle}, J.~E. and {Bate}, M.~R.},
        title = "{Simulations of the grand design galaxy M51: a case study for analysing tidally induced spiral structure}",
      journal = {\mnras},
         year = 2010,
        month = apr,
       volume = {403},
       number = {2},
        pages = {625-645},
          doi = {10.1111/j.1365-2966.2009.16161.x},
archivePrefix = {arXiv},
       eprint = {0912.1201},
 primaryClass = {astro-ph.GA},
       adsurl = {https://ui.adsabs.harvard.edu/abs/2010MNRAS.403..625D}
}

@ARTICLE{2023MNRAS.521.2719Y,
       author = {{Yu}, Haiyang and {Zhu}, Ming and {Xu}, Jin-Long and {Ai}, Mei and {Jiang}, Peng and {Yang}, Yanbin},
        title = "{High-sensitivity H I image of diffuse gas and new tidal features in M51 observed by FAST}",
      journal = {\mnras},
         year = 2023,
        month = may,
       volume = {521},
       number = {2},
        pages = {2719-2728},
          doi = {10.1093/mnras/stad436},
archivePrefix = {arXiv},
       eprint = {2302.03270},
 primaryClass = {astro-ph.GA},
       adsurl = {https://ui.adsabs.harvard.edu/abs/2023MNRAS.521.2719Y}
}

@ARTICLE{2025ApJ...993L..55A,
       author = {{Anand}, Gagandeep S. and {Ben{\'\i}tez-Llambay}, Alejandro and {Beaton}, Rachael and {Fox}, Andrew J. and {Navarro}, Julio F. and {D'Onghia}, Elena},
        title = "{The First RELHIC? Cloud-9 is a Starless Gas Cloud}",
      journal = {\apjl},
         year = 2025,
        month = nov,
       volume = {993},
       number = {2},
          eid = {L55},
        pages = {L55},
          doi = {10.3847/2041-8213/ae1584},
archivePrefix = {arXiv},
       eprint = {2508.20157},
 primaryClass = {astro-ph.GA},
       adsurl = {https://ui.adsabs.harvard.edu/abs/2025ApJ...993L..55A}
}
\begin{appendix}

\section{Physical modelling of RELHICs}\label{sec:relhic_model}

We modelled RELHICs by treating them as spherical gaseous systems in hydrostatic equilibrium embedded within a Navarro--Frenk--White (NFW) dark matter halo. This model enables us to estimate the gas mass bound to a halo of a given virial mass. We assume that the gravitational potential is dominated by the dark matter, with the self-gravity of the gas being negligible.

For a state of hydrostatic equilibrium, the gas density profile can be solved via
\begin{equation}
\frac{1}{\rho}\frac{\mathrm{d}P}{\mathrm{d}r} = -\frac{G M(r)}{r^2},
\end{equation}
where $M(r)$ is the enclosed mass. The pressure--density relation is defined by the $n_{\mathrm{H}}$--$T$ relation followed by gas particles in RELHICs, as described by \citet{Benitez-Llambay2017}. In this framework, the physical properties of a RELHIC are determined by the gas temperature, the halo virial mass ($M_{200}$), and the concentration parameter ($c$).

The gas temperature is coupled to the density through the assumption of thermal equilibrium with the UVB radiation field from \citet{2001cghr.confE..64H}. The hydrostatic equilibrium equations were solved by applying a boundary condition in which the gas pressure at large radii matches the pressure of the intergalactic medium (IGM) at the mean density of the Universe.

In this regime, hydrogen is primarily ionised by the cosmic UVB unless the gas density is sufficiently high for self-shielding. We used the prescription of \citet{Rahmati2013} to compute the ionisation equilibrium and derive the neutral hydrogen (\HI) fraction, from which the \HI\ mass and radial distribution were obtained.

The observable \HI\ mass in the RELHIC framework depends primarily on the halo virial mass ($M_{200}$) and concentration ($c$). For $M_{200}\lesssim10^9\,M_\odot$, the gas remains almost completely ionised by the UVB and becomes undetectable in \HI. Conversely, for $M_{200}\gtrsim5\times10^9\,M_\odot$, gas cooling is expected to become efficient, likely leading to star formation. Consequently, the predicted \HI\ mass rises steeply within the relatively narrow halo mass range of $10^9$--$5\times10^9\,M_\odot$.

\begin{figure}
    \centering
    \includegraphics[width = \linewidth]{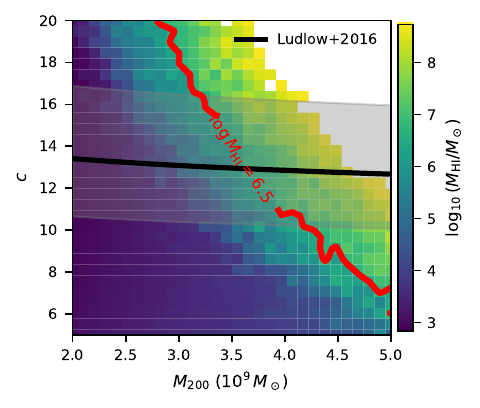}
    \caption{Predicted neutral hydrogen mass in the $M_{200}$--$c$ plane for hydrostatic gas confined within dark matter halos exposed to the ultraviolet background. The colour scale shows $\log(M_{\rm HI}/M_\odot)$, while the red contour corresponds to $M_{\rm HI}=10^{6.5} \,M_\odot$, matching the observed \HI mass of the candidate cloud. The black solid line indicates the median mass--concentration relation from Ludlow et al. (2016), and the grey shaded region shows the corresponding $1\sigma$ intrinsic scatter. Regions where the total gas mass exceeds the cosmic baryon fraction limit, $M_{\rm gas}/M_{200}>0.157$, are excluded. The intersection between the $M_{\rm HI}=10^{6.5}\,M_\odot$ contour and the allowed concentration range constrains the halo mass required to reproduce the observed \HI content.}
    \label{fig:mc}
\end{figure}

Figure~\ref{fig:mc} illustrates the predicted \HI\ mass in the $M_{200}$--$c$ plane. The red contour marks models with $M_{\rm HI}=10^{6.5}\,M_\odot$, corresponding to the observed \HI\ mass of Cloud~N and Cloud~S. The black solid line shows the median mass--concentration relation from \citep{2016MNRAS.460.1214L}, while the grey shaded region indicates the intrinsic $1\sigma$ scatter, corresponding to $10.4<c<16.1$ near $M_{200}\sim4\times10^9\,M_\odot$.

The intersection between the observed-\HI contour and the allowed concentration range constrains the halo mass to $M_{200}\simeq3.7\pm0.4\times10^9\,M_\odot$. Therefore, the inferred halo mass remains confined to a relatively narrow interval within the expected $\Lambda$CDM concentration scatter. Throughout this work we adopt the median concentration, $c=12.8$, as the fiducial value when presenting model profiles.

\section{Validation of mass estimation and kinematics using TNG50 simulations}
\label{sec:tng_validation}
To validate our halo mass estimation method, we used the TNG50 cosmological simulation \citep{Nelson_2019, Pillepich_2019}. We selected a sample of 166 subhalos based on three criteria:
(i) an \HI\ mass between $10^4~\mathrm{M_\odot}$ and $10^7~\mathrm{M_\odot}$,
(ii) zero stellar mass ($M_\star = 0$), and
(iii) being a primary (central) halo to ensure $M_{200}$ is well-defined.

\begin{figure}
    \centering
    \includegraphics[width =\linewidth]{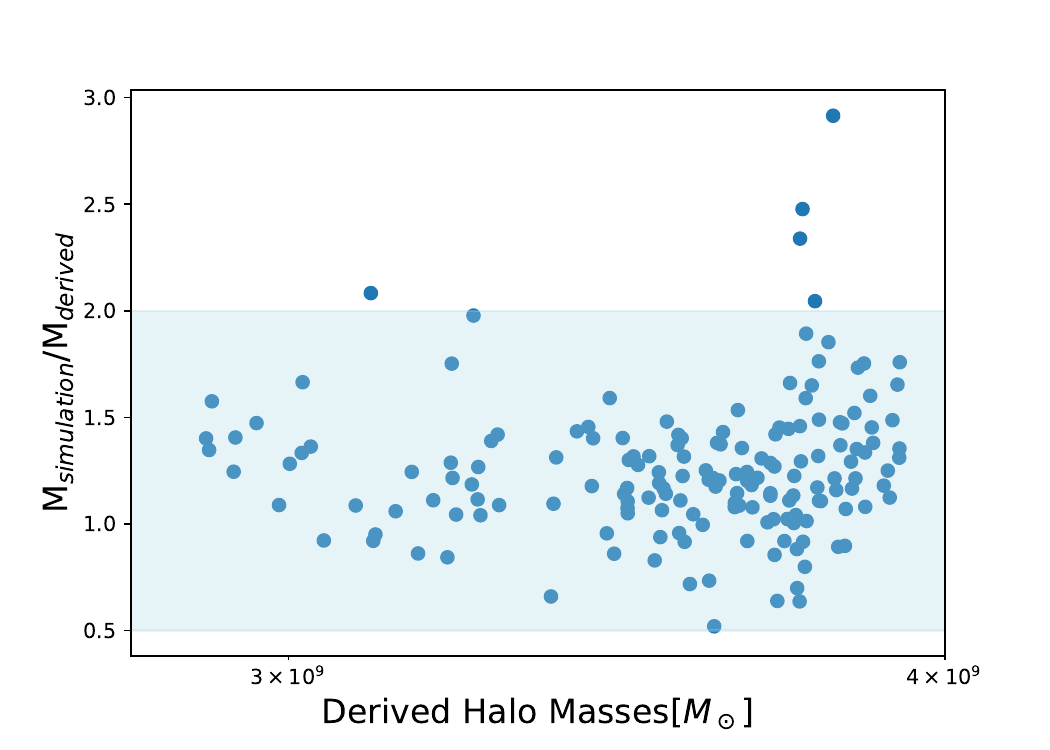}
    \caption{Validation of the RELHIC mass estimation method using the TNG50 simulation. Scatter points represent the selected starless subhalos from the TNG50-1 snapshot. The horizontal axis indicates the halo mass derived from the simulated \HI\ profiles using our RELHIC modelling framework ($M_{\rm derived}$), while the vertical axis shows the ratio of the true subhalo mass from the simulation to this estimated mass ($M_{\rm simulation} / M_{\rm derived}$). The horizontal blue band marks the [0.5, 2.0] ratio range, within which the majority of the sample resides. This agreement within a factor of two demonstrates the robustness of the RELHIC model in recovering total halo masses from \HI\ observations.}
    \label{fig:MSMD}
\end{figure}

The selected halos span a mass range of $10^9$ to $10^{10}~\mathrm{M_\odot}$. Applying our RELHIC-based modelling to the \HI\ masses of these simulated halos, we compared the derived masses with the true $M_{200}$ values from the simulation. As shown in Fig.~\ref{fig:MSMD}, the mass estimates generally agree within a factor of two. This consistency underscores the reliability of our modelling approach in inferring halo masses from starless \HI\ observations.

To examine whether the velocity gradient observed in Cloud~N is compatible with a cosmological origin, we further investigated the internal gas kinematics of the same sample of 166 starless \HI-bearing dark matter halos.

For each halo, we generated synthetic \HI\ data cubes using the \texttt{martini} package \citep{2024JOSS....9.6860O} and convolved them with a Gaussian beam matching the angular resolution of FAST. The mock observations were analysed using the same procedure as applied to the observational data.

\begin{figure}
    \centering
    \includegraphics[width =\linewidth]{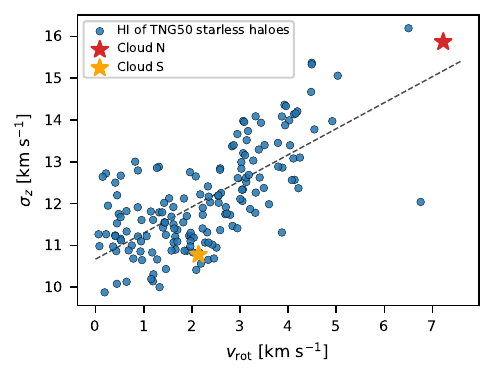}
    \caption{
    Comparison of the simulated starless \HI-bearing dark matter halos with Cloud N and Cloud S. The apparent rotation velocity ($v_{\rm rot}$) is measured after convolution with the FAST beam, while $\sigma_z$ is derived from the second-moment map. Blue dots represent the simulated halos, the red and orange stars denote the two observed clouds, and the black dashed line shows a linear fit excluding the two outliers with $v_{\rm rot}>10~{\rm km\,s^{-1}}$.
    }
    \label{fig:cmp}
\end{figure}

We measured the apparent rotation velocity, $v_{\rm rot}$, from the beam-convolved velocity field. Specifically, we first determined the \HI\ centroid from the moment-0 map and identified the direction with the largest velocity contrast in the moment-1 map. We then divided the cloud into two halves across the centroid and defined
$$
v_{\rm rot}=\frac{1}{2}|v_1-v_2|,
$$
where $v_1$ and $v_2$ are the flux-weighted mean line-of-sight velocities of the two halves. The velocity dispersion, $\sigma_z$, was calculated from the second-moment map within the same \HI\ mask. Since the intrinsic shapes and inclinations of the clouds are unresolved by FAST, $v_{\rm rot}$ should be interpreted as an apparent, beam-smoothed measure of ordered motion rather than a true circular velocity. To assess whether the observed kinematics are consistent with a cosmological origin, we compare Cloud N and Cloud S with simulated starless \HI-bearing dark matter halos from TNG50 in Figure~\ref{fig:cmp}.

We find that the simulated starless \HI-bearing dark matter halos are not purely pressure-supported systems. Although thermal pressure provides the dominant support, many halos exhibit measurable ordered gas motions that appear as velocity gradients in the beam-convolved mock observations. Both Cloud~N and Cloud~S fall within the range spanned by the simulated halo sample in the $v_{\rm rot}$--$\sigma_z$ plane, with Cloud~N lying toward the high-$v_{\rm rot}$ end and Cloud~S occupying the low-$v_{\rm rot}$ regime. This comparison indicates that the observed difference in their internal kinematics is not inconsistent with a cosmologically formed population of starless \HI-bearing dark matter halos.

\section{Detailed \HI\ moment maps and kinematics}

\begin{figure*}
        \includegraphics[width = \linewidth]{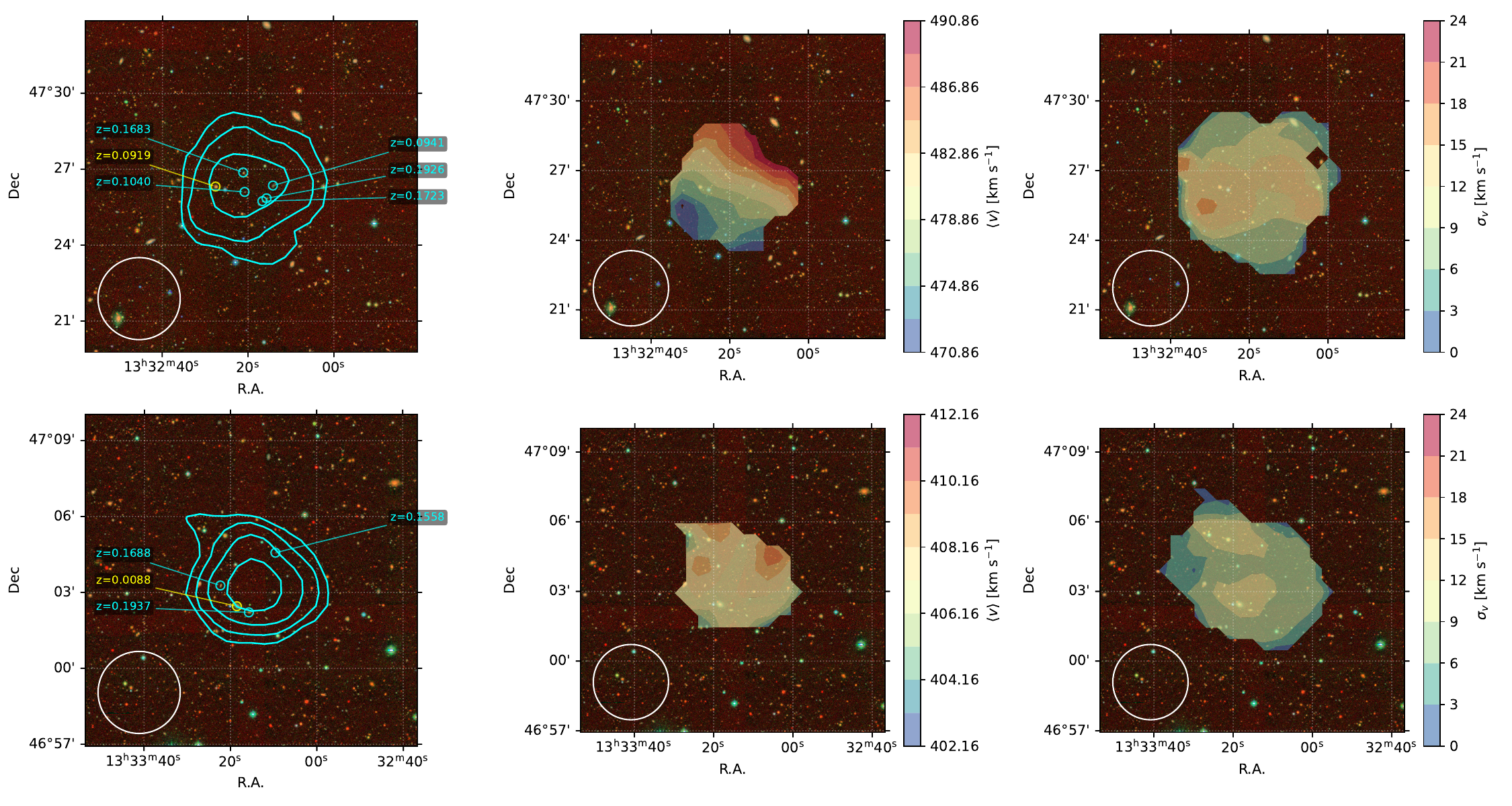}
    \caption{Detailed \HI\ moment maps for the RELHIC candidates Cloud~N (upper panels) and Cloud~S (lower panels) overlaid on a Legacy Survey optical image. From left to right, the columns show the \HI\ column density (moment~0), where the contours correspond to column density levels of $10^{17.75}$, $10^{18}$, $10^{18.25}$, $10^{18.5}$, and $10^{18.75}\,\mathrm{cm^{-2}}$ from outer to inner regions; the intensity-weighted line-of-sight velocity field (moment~1); and the velocity dispersion map (moment~2). The white circles in each panel indicate the FWHM beam size of the FAST telescope. The extragalactic sources that fall within one FWHM beam size are also marked in the moment~0 map. Cyan symbols represent the photometric redshifts and yellow symbols represent the spectroscopic redshifts.}
    \label{fig:zoom in}
\end{figure*}

\begin{figure*}
        \includegraphics[width=\textwidth]{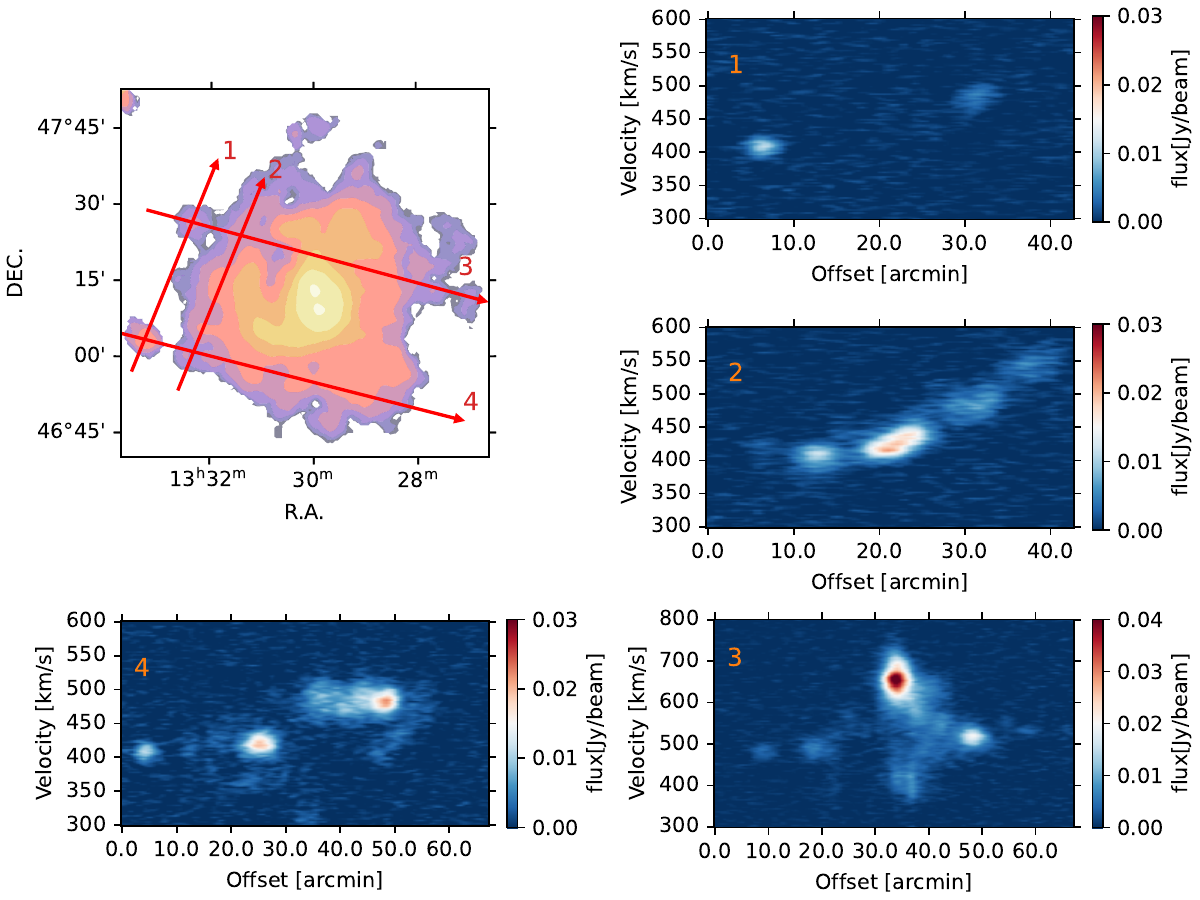}
    \caption{P-V diagrams for the RELHIC candidates along several designated slices. Path~1 connects Cloud~S and Cloud~N, while path~2 provides a parallel reference trajectory. Paths~3 and 4 illustrate the velocity connection between Cloud~S, Cloud~N, and the ambient \HI\ gas associated with the host galaxy, M51. The kinematic continuity observed along these paths strongly suggests that both clouds are physically associated with the M51 system rather than being foreground or background objects.}
    \label{fig:PV1}
\end{figure*}
The moment maps (0, 1, and 2) for both clouds are presented in Fig.~\ref{fig:zoom in}, overlaid on deep optical imaging of the region.
Both clouds reach peak \HI\ column densities above $10^{18.5}\,\mathrm{cm^{-2}}$.
Cloud~N shows a clearer velocity gradient, while Cloud~S shows weaker ordered motion and does not display a visually obvious rotational pattern in the moment-1 map.
The velocity dispersion, as indicated by the moment~2 map ($\sigma$), is approximately $20\,\mathrm{km~s^{-1}}$ for both clouds.

In Fig.~\ref{fig:zoom in}, we mark all extragalactic sources from the DESI Legacy Imaging Surveys Tractor Catalog that fall within one FWHM beam size of each \HI\ centroid, along with their available spectroscopic or photometric redshifts.
Within the beam centred on Cloud~N, we identify one source with a photometric redshift and five sources with spectroscopic redshifts.
Within the beam centred on Cloud~S, we identify one source with a spectroscopic redshift and three sources with photometric redshifts.
In all cases, the source redshifts are substantially larger than that of M51 ($z \approx 0.0013$, corresponding to $\sim$460~km~s$^{-1}$), confirming that none of these sources is a viable optical counterpart to the \HI\ clouds.

The position-velocity (P-V) diagrams shown in Fig.~\ref{fig:PV1} reveal that the radial velocities of the candidates are kinematically consistent with the M51 system environment, suggesting that they are gravitationally associated with, or at least located at the same distance as, M51.
This kinematic consistency justifies our assumption of an 8~Mpc distance for both clouds in subsequent mass calculations.
\end{appendix}  
\end{document}